# Application-Layer FEC Scheme Configuration Optimization via Hybrid Simulated Annealing

César Díaz, Julián Cabrera, Fernando Jaureguizar and Narciso García

*Abstract*—An optimization technique based on an adapted combination of simulated annealing (SA) and tabu search (TS) is presented. This method aims at finding near-optimal unequal error protection (UEP) application-layer FEC (AL-FEC) code configurations. This approach is intended to smartly protect audio and video transmission over IP networks when hard time restrictions apply. The considered code is a UEP version of the widely-used Pro-MPEG COP3 codes enabling the use of several matrices of dissimilar size and thus of unequal recovery capability. Finding the optimal configuration frequently requires the evaluation of a large solution space. So, to fulfill the imposed constraints, SA is adapted to the specifics of the scenario. In particular, the annealing schedule is conditioned by the real-time restrictions. Furthermore, solution neighborhood structures are determined by a proposed definition of distance between protection configurations, which, jointly with TS, conditions the selection of candidate solutions. Experimental results show a significantly improved performance of the optimization process, which invariably fulfills imposed timing constraints, at the expense of a very low distortion increase, when compared to using exhaustive search. These results allow the use of UEP Pro-MPEG COP3 codes for protecting video and audio transmission, which distinctly outperforms the standard code in a wide range of scenarios[1].

*Index Terms*—Forward Error Protection, Optimal Resource Allocation, Real Time, Simulated Annealing, Tabu Search, Unequal Error Protection, Video and Audio Streaming.

## I. Introduction

In scenarios where video- and audio-related services are provided through IP networks (e.g. television broadcasting, video on demand (VoD), IP television (IPTV), voice over IP or videoconferencing), application-layer forward error correction (AL-FEC) techniques are commonly used to increase the reliability of the communication channel, and so ensure to a large extent a target quality [1], [2]. However, many of these services, particularly those that involve live event broadcasting or some kind of

[1] This work has been partially supported by the Ministerio de Economía, Industria y Competitividad of the Spanish Government under project TEC2013-48453 (MR-UHDTV), and the Spanish Administration agency CDTI under project IDI-20150950 (Tranvideoadap).

César Díaz is with the Grupo de Tratamiento de Imágenes (GTI), Information Processing and Telecommunications Center (IPTC) and ETSI Telecomunicación, Universidad Politécnica de Madrid, 28040 Madrid, Spain (e-mail: cdm@gti.ssr.upm.es).
Julián Cabrera is with the Grupo de Tratamiento de Imágenes (GTI), Information Processing and Telecommunications Center (IPTC) and ETSI Telecomunicación, Universidad Politécnica de Madrid, 28040 Madrid, Spain (e-mail: Julian.Cabrera@gti.ssr.upm.es).
Fernando Jaureguizar is with the Grupo de Tratamiento de Imágenes (GTI), Information Processing and Telecommunications Center (IPTC) and ETSI Telecomunicación, Universidad Politécnica de Madrid, 28040 Madrid, Spain (e-mail: fjn@gti.ssr.upm.es).
Narciso García is with the Grupo de Tratamiento de Imágenes (GTI), Information Processing and Telecommunications Center (IPTC) and ETSI Telecomunicación, Universidad Politécnica de Madrid, 28040 Madrid, Spain (e-mail: narciso@gti.ssr.upm.es).

interactivity, are not only demanding in terms of quality, but also of latency. In these cases, the selected protection mechanisms must perform in real time [3].

Furthermore, as a result of the encoding process, different parts of the packet stream are of unequal importance, because of the dissimilar impact of their potential loss on the quality of the content presented to end users, due to error propagation. Hence, unequal error protection (UEP) schemes are frequently utilized to smartly allocate available resources among source data in regard to their importance [4]. In the above mentioned scenarios, the resources to allocate are the channel coding rate, but they could also be the transmit power, the modulation mode, etc. [5].

In some UEP schemes, the distribution of resources responds to an a priori arrangement. In these cases, a pre-established fixed categorization of data and a rather constant distribution of data among these categories are habitually assumed [6], [7]. However, the distribution of resources is usually the result of an optimization problem, in which the optimal distribution is one in a set of feasible solutions. This problem is tackled differently regarding the characteristics of the protection mechanism (e.g. whether the distribution of resources is set before the transmission begins -a single instance that is solved offline-, or periodically along the course of the transmission -a series of instances that are solved on the fly), the level at which the stream is analyzed and at which resources are distributed (e.g., macroblock, slice, frame, layer, etc.), and the imposed restrictions (in particular regarding latency and computation complexity).

Regarding combinatorial optimization, strategies that work with rather small solution spaces or are not conditioned by strict restrictions, usually perform an exhaustive search to find the optimal solution [8], [9]. On the other hand, if the optimization problem fulfills the necessary requirements (e.g. continuous first partial derivatives, convexity, etc.), or a relaxation method can be applied, strategies commonly turn to exact optimization algorithms to find the optimal solution (e.g. integer or mixed-integer programming). In particular, there exist numerous proposals posing optimization problems that can be formulated in terms of rate-distortion. Usually, in these strategies, channel coding is seen from a high level of abstraction and they do not aim at a further optimization of the code. Specifically, designers frequently chose maximum-distance-separable (MDS) channel codes (e.g. Reed-Solomon codes), that is, ideal codes, to that end. In these cases, the optimization problem is usually solved by the method of Lagrange multipliers [10], [11].

However, in many situations, protection schemes cannot include exact optimization algorithms to solve the optimization problem. This is typically the case when a white-box perspective is assumed, in which the designer aims at obtaining the best internal parameters values of the protection mechanism to optimally distribute the available resources. In addition, regarding combinatorial optimization, the characteristics of the scenario may not allow the use of brute-force search. In these circumstances, many proposals use iterative methods to find the optimal solution. The iterative methods can be either standard (e.g. iterative linear programming [12], branch and bound method [13]) or created ad hoc [14], [15].

Finally, in scenarios limited by hard restrictions, either regarding latency or computation complexity, with a large solution space, where neither exact optimization algorithms nor iterative methods can be used, metaheuristics are commonly employed to guide the search process [16]. These approaches obtain sufficiently good solutions, that is, solutions that may not be optimal but are considered good enough for the purposes of the problem, given the imposed restrictions. In this regard, designers mainly opt for search methods based on genetic algorithms (GA) [17], [18] or guided local search (GLS) [19], [20]. In principle, a broad set of metaheuristics are available to designers [21], from population-based nature-inspired approaches like evolutionary algorithms (e.g. the above mentioned GA [22]) or swarm algorithms (e.g. ant colony optimization –ACO- [23] or particle swarm optimization –PSO- [24]), to single-solution options like simulated annealing (SA) [25], tabu seach (TS) [26], or the also already mentioned GLS [27]. Nevertheless, the selection of a metaheuristic highly depends on the actual context at hand, as its performance relies on how its characteristics fit in with those of the scenario and the imposed limitations. In the resource allocation problem considered in this paper, the time that takes the strategy to reach acceptable solutions and the simplicity of the scheme are of particular concern. The former is important so as to comply with real-time application latency requirements, whereas the latter responds to practical reasons, as the employed algorithm is intended to be incorporated in transmission modules to be deployed in a potentially high range of devices, from high-level computational capacity servers to computationally limited terminals. Of the available approaches, SA is one of the options that is better suited for the scenario: it can meet such rigid conditions regarding time and complexity, whereas still providing appropriate solutions. Although, depending on the context, it may be outperformed by other algorithms in the long run regarding how close final solutions are to the overall optimal one, contrary to other metaheuristics, SA is well known for being a "quick starter", that is, a method that is able to obtain good enough solutions in short periods of time [28]-[30]. Moreover, it is highly prone to problem-specific adaptations. These are the reasons why this rather simple metaheuristic has been successfully widely applied to solve many combinatorial problems within and outside the area of communications [31]-[33]. In addition, with the aim of making this memoryless technique more effective, it can be hybridized with TS, which provides memory structures to enhance the search of solutions [34], [35]. TS encourages intensification or diversification upon convenience, and so helps guide the search to reach better solutions in very short time periods.

Consequently, a fast and robust optimization technique based on SA and backed by TS is proposed in this paper. This technique aims at finding near-optimal AL-FEC code configurations when strict time restrictions apply. The considered channel code is a new UEP version, first proposed in [36], of the broadly-used Pro-MPEG COP3 AL-FEC codes introduced by the Pro-MPEG Forum in its Code of Practice 3 r2 [37]. This new version enables the use of several matrices of dissimilar size per protection block, in such a way that unequal code rates can be applied to different groups of data packets in regard to their importance, without increasing the amount of devoted resources. In the considered protection strategy, the optimal configuration (i.e. the number of matrices and their size that minimize the overall expected distortion) is computed periodically, so that it adapts to the varying behavior of both the packet stream and the communication channel. Moreover, the number of feasible configurations can be very high, as it grows along with the number of data and repair packets in the block. So, the described combinatorial problem, with a potentially large solution space, and very restricted processing time, is solved by the proposed hybrid procedure. Moreover, although adapted to the mentioned scheme, the proposed procedure can be straightforwardly adjusted to suit the characteristics of other UEP schemes.

The rest of the paper is organized as follows. Section II presents a quick description of important characteristics of the video stream, key for the use and put in practice of the adopted UEP AL-FEC codes. Section III includes a detailed description of the considered protection codes. In Section IV, the problem description is introduced. Section V presents in depth the proposed hybrid SA metaheuristic. Experiments and results are presented in Section 0. Finally, Section VII, includes the conclusions of the paper.

## II. SOURCE CODING AND PACKETIZATION

In this paper, it is assumed that the video sequence is encoded using any video coding standard that enables the subsequent identification and assessment of the different parts of the resulting bit stream. For instance, the latest video coding standards, MPEG-4 Part 10 H.264/AVC [38] and MPEG-H Part 2 H.265/HEVC [39], use a Network Abstraction Layer (NAL) to organize the resulting encoded bitstream into self-contained units (NAL units) headed by a set of bytes presenting information on the data included in that unit. The header of these units can be later on accessed to check basic features of the data included in that unit, such as the type of slice or its position along the stream, key for assessing the relative importance of the data.

In addition, given the characteristics of the scenario, where the service to be provided is conditioned by strict time restrictions, it is assumed that either the broadly used Real-time Transport Protocol (RTP) [40] or the new MPEG Media Transport Protocol (MMTP) [41] is employed to carry the encoded video data. The organized encoded video data can be mapped directly into the transport protocol packets or, in the case of RTP, possibly through an intermediate encapsulation step (e.g. MPEG-2 Transport Stream –MPEG-2 TS- [42]). In this way, thanks to the information included in the transport protocol packet, the unit and any intermediate encapsulation headers, it is possible to assess the importance of each packet through estimating the distortion that would be introduced in the decoded video sequence presented to users if the packet is lost.

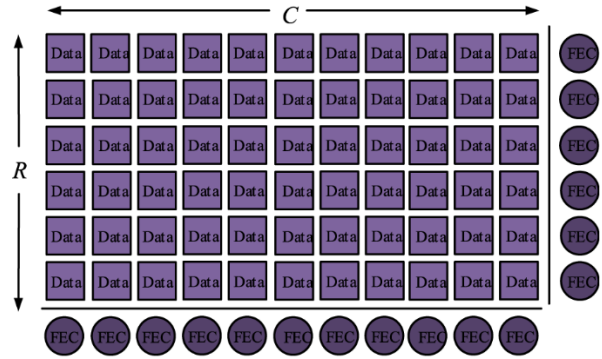

**Fig. 1. Standard Pro-MPEG COP3 codes: all data packets in the same protection block are arranged in one matrix. Repair packets might be generated row- and column-wise**

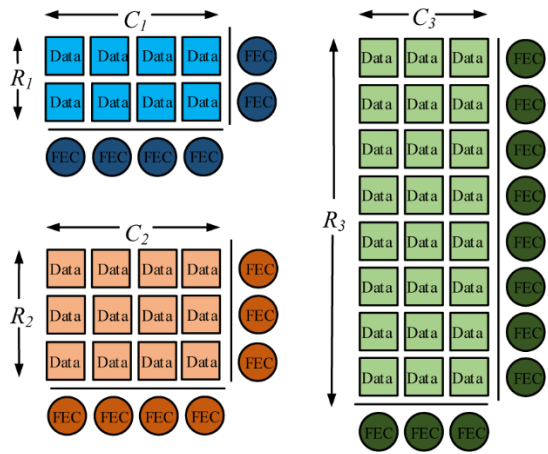

**Fig. 2. UEP Pro-MPEG COP3 codes. Example with three matrices. All matrices may have different dimensions, hence offering different degrees of protection**

### III. UEP PRO-MPEG COP3 CODES

*A. Standard codes*

*1) Introduction*

The Pro-MPEG COP3 AL-FEC codes were introduced by the Pro-MPEG Forum in its Code of Practice 3 r2 [37] and were later on standardized by the SMPTE in its specification 2022-1 [43]. Since then, they have been included in a number of standardization documents related to IPTV and video broadcasting, including the DVB standard for IPTV (ETSI TS 102 034 [44]) and documents of various organizations: ATIS, ETSI, ITU-T, the Open IPTV Forum (who transferred its technical activities to the HbbTV Association in 2014) [45], etc.. In the most recent standards (e.g. MMTP), they have been considered, like other codes, under the FEC Framework, a protection mechanism abstraction layer that enables a more generic and easier integration of FEC codes.

These codes have been widely deployed, both alone and jointly with a second code (e.g. Raptor codes [46]), due to their very appealing features. Particularly, their low complexity and their capability to cope with burst errors are very well suited for scenarios involving real-time video streaming over packet-switch networks.

The Pro-MPEG COP3 codes are erasure codes that operate at application layer. At this level, networks behave like packet erasure channels, where error occurrences: (i) affect whole network packets; (ii) never result in the reception of corrupted packets (packets turn out either absolutely unaltered or lost); and (iii) the location of the lost packets is known. Error events of this kind include the discarding of packets at intermediate routers (due to network congestion), transmission timeouts (an excessively delayed packet is in practice a lost packet to many applications), or the detection and subsequent rejection of corrupted packets (i.e., containing bit errors not corrected at lower levels). For that reason, Pro-MPEG COP3 codes commonly, and this is also the case in this paper, use whole network packets (e.g. RTP- or MMTP-wrapped) as source symbols [47]. To ensure that all symbols are of the same length, padding bytes are used when needed. Furthermore, these codes are systematic. This means that source symbols are embedded unaltered in the encoder output, that is, codewords are made up of the original data packets and of the repair packets that result from the encoding process. All the generated redundancy is then allocated in the repair packets. The original data packets and the generated repair packets are sent in two different flows.

2) Coding and decoding processes

In the standard Pro-MPEG COP3 codes, data packets are organized in matrices of $R$ rows and $C$ columns. Packets are arranged in sequence number order, filling the matrix row by row, from the column most to the left to the one most to the right. Protection packets can then be generated both row-wise and column-wise by XORing the associated data packets, bit by bit, as illustrated in Fig. 1. So, the number of data packets per FEC block equals $R \cdot C$, whereas the number of repair packets depends on whether one or both dimensions are employed: $R$, if only the row dimension is used, $C$, if only the column dimension is employed, and $R + C$, if both dimensions are used. The particularity of these codes is that they do not behave as an usual $(n, k)$ one, where $k$ is the number of data packets in the block (the matrix, in this case) and $n$ this same number plus that of generated repair packets, but as an aggregation of inner codes, where each row and column acts independently. Indeed, regarding row-wise generated redundancy, each row constitutes an $(C + 1, C)$ code and, and regarding column-wise generated repair packets, each column is basically a $(R + 1, R)$ code. So, for each row/column, protection packets can help if only one of the source packets in the row/column is lost. Therefore, the parity packets generated row-wise are suited to deal with independent packet losses, whereas column-wise-created FEC packets can deal with burst errors, as they are basically the result of applying an interleaving step of depth $C$ prior to performing the XOR operations. The reason is that any error burst leading to losing two or more data packets in the same row directly overcomes the recovery capability of the repair packet associated with that row. However, this same burst, if it leads to the loss of *C* of less data packets and no other bursts occur, can be coped with using the repair packets generated column-wise, as it will result in just one lost packet per protection packet.

The procedure to recover lost packets depends on which dimensions are used to generate FEC packets. The recovery algorithm loops through the corresponding dimension of the matrix, selecting those columns and/or rows where only one data packet is lost and the repair packet has been received. The available data and repair packets of these columns or rows are XORed, resulting in the reconstruction of the lost packet. If both dimensions are used, loops are carried out sequentially so that recently rebuilt data packets can be used for further packet recovery.

Additionally, in the considered scenario, packets are sent sequentially, that is, repair packets are sent right after the last data packet belonging to the matrix is transmitted. The reason is that such sending arrangement allows maximum separation between any two packets belonging to the same column, either data or repair, thus fostering interleaving and so error decorrelation in case of error burst.

*3)* *Performance*

The Pro-MPEG COP3 codes perform very effectively for sufficiently low channel packet loss rates (PLRs). However, they become less successful as this value increases, thus potentially leading to intolerable levels of degradation of the video presented to final users [3].

*B.* *Proposed UEP version*

With the aim of improving the performance of standard Pro-MPEG COP3 codes at high PLRs, a packet-level UEP version was proposed [36]. This approach basically enables the use of a number of matrices of unequal size per protection block, as shown in Fig. 2, without devoting more resources per block than the standard case. In this way, unequal code rates can be applied to different groups of data packets in regard to their unequal importance in terms of the degradation that their loss may cause in the decoded sequence. As a result of applying this unequal protection scheme, the Pro-MPEG COP3 codes can significantly improve their performance, leading to acceptable levels of video quality, even if the PLR remains high. The reason is that, whilst in the standard case this rate is roughly uniform for all types of data packets, when applying the UEP version, the resulting PLR is lower for more important packets and greater for less important ones. A more detailed description of this approach, as well as a study of its performance, can be found in the experiments and results section and in [36].

IV. **PROBLEM DESCRIPTION**

A. *Optimization problem formulation*

The problem is formalized in the following terms. Let $N_M$ be the number of matrices in a configuration, and $C_m$ and $R_m$ respectively the number of columns and rows in matrix $m$, $1 \leq m \leq N_M$. It is assumed that matrix 1 protects the $C_1 \cdot R_1$ most relevant packets in the protection block, matrix 2 the following $C_2 \cdot R_2$, and so on.

The relevance of the different packets is the result of the application of a particular distortion model to the bit stream (e.g. [7]-[15]). The specific distortion model that is used is out of the scope of this paper. However, it is assumed that the resulting values of that process are known by the protection system.

The goal of the optimization problem is to find the most convenient protection configuration, $s^{opt}$, that is, the combination of values of the variables $N_M, C_1, R_1, \ldots, C_{N_M}, R_{N_M}$ that minimizes the overall expected distortion introduced by the $N_P$ data packets in the block, $D_T$. The general formulation of the problem is as follows:

$$\min_{N_M, C_1, R_1, \ldots, C_{N_M}, R_{N_M}} D_T = \sum_{p=1}^{N_P} D_p \cdot P_p \qquad (1)$$

where $D_p$ is the distortion associated with losing packet $p$, which, as mentioned, results from applying the distortion model, and $P_p$ is the likelihood of losing packet $p$. $P_p$ not only depends on the behavior of the channel, but also on the selected protection configuration, and, more specifically, on the protection matrix where packet $p$ is allocated.

The problem is subject to the following conditions:

- All data packets have to be protected, and every data packet is protected by one and only one matrix. Moreover, only the last matrix, matrix $N_M$, may need padding packets to complete its packet distribution:

$$\sum_{m=1}^{N_M-1} C_m \cdot R_m + C_{N_M} \cdot (R_{N_M} - 1) < N_P \leq \sum_{m=1}^{N_M} C_m \cdot R_m \qquad (2)$$

- Assuming an essentially bursty channel, parity packets are generated only column-wise:

$$\sum_{m=1}^{N_\text{M}} C_m = N_\text{FEC} \tag{3}$$

where $N_\text{FEC} = N_\text{P} \cdot ((1/r_\text{FEC})-1)$ is the number of repair packets, and $r_\text{FEC}$ is the imposed minimum code rate.

Due to the imposed time limitations, only solutions considering 1 to $N_{\text{M}_\text{max}}$ matrices will be able to be tested, and so $N_\text{M} \in \{1, \dots, N_{\text{M}_\text{max}}\}$. The overall optimization problem is then divided into $N_{\text{M}_\text{max}}$ subproblems. Each subproblem works with configurations of different number of matrices and delivers the best solution that it has been able to find. So, at the end, $N_{\text{M}_\text{max}}$ candidate solutions to the overall optimization problem will be available: $s_1^\text{opt}$ to $s_{N_{\text{M}_\text{max}}}^\text{opt}$. The overall optimal solution, $s^\text{opt}$, will be the best one among them.

Furthermore, assumed the imposed conditions, it can be seen that each subproblem has a number of degrees of freedom that depends on the number of matrices of the configurations that it works with. In particular, if the configurations are made up of $N_\text{M}$ matrices, the problem will have $2 \cdot (N_\text{M} - 1)$ degrees of freedom. This calculation considers the dimensions of all the matrices but one, whose size can be computed once the others have been set.

B.  *Solution space of the subproblem with configurations of $N_\text{M}$ matrices*

Let $\Omega_{N_\text{M}}$ be the solution space of the subproblem, i.e., the number of possible combinations of values that fulfill the expressed restrictions, if $N_\text{M}$ matrices are employed. Then, for given values of variables $N_\text{P}$ and $N_\text{FEC}$, the cardinality of the solution space, $N_{\Omega_{N_\text{M}}}$, can be computed through the following iterative and recursive equation:

$$N_{\Omega_{N_\text{M}}}(N_\text{P}, N_\text{FEC}) = \sum_{i=1}^{N_\text{FEC}-N_\text{M}+1} \sum_{j=1}^{\left\lfloor \frac{N_\text{P}-N_\text{FEC}+i}{i} \right\rfloor} N_{\Omega_{N_\text{M}-1}}(N_\text{P} - i \cdot j, N_\text{FEC} - i) \tag{4}$$

where the base case of the recursive equation:

$$N_{\Omega_1}(N_\text{P}, N_\text{FEC}) = 1 \tag{5}$$

Each iteration is used to set the number of columns and rows of the current matrix, and each recursion step considers all the feasible solutions once the dimensions of the current matrix are known. So, in the first recursion step, each iteration sets the values of $i$ and $j$, which are the number of columns and rows of the first of the $N_\text{M}$ matrices, respectively. In this way, two variables out of the $2 \cdot (N_M - 1)$ ones are set. The expression inside the summations then considers all the feasible combinations of sizes of the remaining $N_M - 1$ matrices, i.e., the feasible combinations of values of the remaining $2 \cdot (N_M - 2)$ variables. In the following recursion step, each iteration sets the values of $i$ and $j$, that is, the size of the second matrix, thus setting two more of the original $2 \cdot (N_M - 1)$ variables. In the same way as in the previous step, the expression inside the summations considers all the feasible configurations, once the dimensions of the first two matrices are set. The recursion continues in the same way. The base case in (5) considers all the feasible combinations, once the size of all matrices but the last one are set. Since all the $2 \cdot (N_M - 1)$ variables are already set, the number of remaining feasible solutions is one.

$N_{\Omega_{N_M}}$ can be quite vast, particularly for high values of $N_P$, $N_{FEC}$ and $N_M$.

## V. PROPOSED OPTIMIZATION PROCEDURE

The pursued goal is, as mentioned before, to select the most suitable protection configuration to minimize the deterioration of the transmitted video or audio caused by packet losses, given a limited extra bitrate budget for this purpose and a maximum time of accomplishment. To that end, an adapted metaheuristic based on the hybridization of two optimization techniques, simulated annealing (SA) and tabu search (TS), is proposed: hybrid simulated annealing (HSA).

The first step of the proposed procedure is to reduce the number of feasible solutions to each of the subproblems. For that purpose, two extra restrictions are included to the optimization problem. These new conditions are consistent with the objective of protecting data packets considering their relevance, providing stronger protection to more important packets.

The second step incorporates the optimization process strictly speaking. It is based on an adapted SA. However, the present procedure enhances this algorithm by including memory structures via TS. Moreover, the proposed HSA strategy is adapted to the specifics of the scenario. In particular, the annealing schedule is conditioned by the restrictions imposed to fulfill real-time requirements, and the neighbors of a solution are determined by a proposed definition of distance between configurations. The TS procedure is used to keep track of previously visited solutions, and help determine which solutions in the solution space can be candidates for the next move. Moreover, it makes the system periodically move back to so-far best solutions.

In the next subsections, both steps are discussed in depth.

### A. Solution space reduction

The solution space is delimited by adding the two next restrictions:
- More resources devoted to more important packets. Using more columns boosts interleaving, and so, error decorrelation, in channels with memory, since data packets in the same column are more distant in the packet stream. Thus, for each feasible solution of $N_M$ matrices, the number of columns of the matrices will be monotonically decreasing as the index of the matrix increases:

$$C_{m_1} \geq C_{m_2}, 1 \leq m_1 \leq m_2 \leq N_M \tag{6}$$

- Lower code rates devoted to more important packets. Using fewer rows reinforces the capability of the repair packets to rebuilt data packets, as the likelihood of losing more than one packet per column decreases. In this case, for each feasible solution of $N_M$ matrices, the number of rows of the matrices will be monotonically increasing as the index of the matrix increases:

$$R_{m_1} \leq R_{m_2}, 1 \leq m_1 \leq m_2 \leq N_M \tag{7}$$

If these extra restrictions are considered, the new number of feasible solutions of a subproblem working with configurations of $N_M$ matrices, $\widetilde{N}_{\Omega_{N_M}}$, can be computed as follows:

$$\widetilde{N}_{\Omega_{N_M}}(N_P, N_{FEC}, C_{max}, R_{min}) = \sum_{i=\left\lceil\frac{N_{FEC}}{N_M}\right\rceil}^{\min(C_{max}, N_{FEC}-N_M+1)} \sum_{j=R_{min}}^{\left\lfloor\frac{N_P-N_{FEC}+i}{i}\right\rfloor} \widetilde{N}_{\Omega_{N_M-1}}(N_P - i \cdot j, N_{FEC} - i, i, j) \tag{8}$$

where the base case of the recursive equation:

$$\widetilde{N}_{\Omega_1}(N_P, N_{FEC}, C_{max}, R_{min}) = \begin{cases} 1, \text{if } \left\lfloor\frac{N_P - N_{FEC}}{N_{FEC}}\right\rfloor \geq R_{min} \\ 0, \text{otherwise} \end{cases} \tag{9}$$

This equation works in the same way as the one in the previous section, only incorporating the new restrictions. In this regard, $C_{max}$ and $R_{min}$, which respectively are the maximum number of columns and the minimum number of rows that the current matrix can have, according to (6) and (7), restrict the range of feasible values of the iterators. So, if the maximum number of rows that the current matrix could have had, regarding the remaining number of data and repair packets, is not higher than the imposed minimum number of rows, that is, if $\left\lfloor\frac{N_P-N_{FEC}+i}{i}\right\rfloor < R_{min}$, then this summation becomes an empty sum, and no further recursion is performed. This condition is also explicitly considered in the base case, i.e., the one that considers all the feasible combinations, once the size of all matrices but the last one are set. Lastly, regarding the first recursive step, i.e., the dimensions of the first matrix, $C_{max}$ equals $N_{FEC} - N_M + 1$ and $R_{min}$ is 1.

Those conditions make the number of feasible solutions go down. In Section 0, the impact of the reduction of the space solution in the optimization procedure is analyzed.

B.     Proposed hybrid simulated annealing (HSA)-based metaheuristic

1)     Base methods

   a)     Simulated annealing

SA is a metaheuristic for providing sufficiently good solutions to optimization problems. In particular, it is broadly utilized to approximate the optimal solution to nonlinear combinatorial optimization problems, where there exists a global minimum among several local minima. It is especially useful for problems with large solution space, limited computation capacity or hard time restrictions. Moreover, it is highly prone to problem-specific adaptations.

It was introduced in 1983 by Kirkpatrick et al. [48] as an emulation of the physical process through which a molten metal is slowly cooled so when the minimum temperature is reached, this happens at a minimum energy configuration.

The basic iteration of the SA procedure involves randomly selecting a feasible solution, $s_n$, out of the set of neighbors of the current solution, $s_c$, and moving to it with some probability. The move to this new solution always takes place if its cost, $D(s_n)$, is lower than that of the current solution, $D(s_c)$, (downhill move) and with a probability lower than 1 if it is higher (uphill move). Thanks to allowing these uphill moves, the heuristic can escape from local minima and progress toward more suitable solutions. The probability of acceptance of a more-costly solution, $P_a$, depends on the difference of cost between the two solutions and on a global control parameter called temperature, $T$, originally and typically as follows:

$$P_a = \exp((D(s_c) - D(s_n))/T) \qquad (10)$$

So, the lower the difference of cost is and the higher the temperature is, the higher it becomes the probability of accepting the new solution.

The algorithm then consists of an outer loop and an inner loop. The outer loop is basically employed to update the temperature. The temperature is initialized to some value, greater than zero, and is decreased at each step following a given annealing schedule.

Every inner loop carries out a series of basic iterations to explore the solution space, as described before. Within each iteration, solution $s_n$ is selected from the ones that make up the neighborhood of solution $s_c$. The neighbors of a solution are the solutions in the solution space fulfilling some conditions. SA is originally a local search method, and so, the neighborhood of a solution is made up of the solutions that are located the closest to it, regarding a given definition of distance. However, when facing large solution spaces and strict time constraints, this approach encounters difficulties in finding the optimal solution, as the solution space cannot be properly inspected. In order to overcome this problem, some authors have proposed to use different, more flexible neighborhood structures [49], [50]. In this paper, an approach based on varying scale neighborhood structures is used. In this way, the definition of neighbor changes throughout the performance of the procedure, benefiting it in regard of the pursued objective at each stage.

*b)    Tabu search*

TS is a metaheuristic search method proposed by F. Glover in 1986 [51] for solving combinatorial optimization problems. It employs flexible memory to avoid being trapped at local minima, either by forbidding or penalizing moves that would return to a recently visited solution.

This procedure generates neighborhoods of the current solution in accordance to a set of rules and banned solutions included in the so-called tabu lists. A neighborhood is used to search for new solutions to move to. If all the solutions in a neighborhood are tested, a new one is utilized. This iterative method continues until a stop condition is met.

The memory structures are divided into three categories: short-, intermediate- and long-term, considering the scope of the rules and banned moves included in the lists:

- Short-term memory structures are used to prevent the algorithm from cycling, that is, from returning to recently visited regions of the solution space. To that end, recently visited solutions are registered in a tabu list.
- Intermediate-term memory structures are employed to implement so-called aspiration criteria. Those criteria are used to intensify the search of solutions in promising regions of the solution space. For that purpose, short-term prohibitions are relaxed by allowing certain moves previously included in the tabu list.
- Long-term memory structures are used for diversification, that is, to promote the search of solutions in new or scarcely explored regions of the solution space. The objective of these structures is then basically the opposite of that of the intermediate-term ones.

```
// Subproblem with N_M equal to 1
N_M ← 1
T^init ← s_1^opt
terminate_procedure ← False
while not terminate_procedure
    // New subproblem
    // Subproblem global parameters
    N_M ← N_M + 1
    F_{Ω_{N_M}} ← [ ]
    update d_{N_M}^init
    // First outer iteration
    T ← T^init
    d_{N_M} ← d_{N_M}^init
    update N(s, d_{N_M})
    s ← random(C(s))
    s^best ← s
    include s in F_{Ω_{N_M}}
    compute I^inner
    // Inner loop
    do inner_loop
    // Next outer iterations
    compute I^outer
    if I^outer < 2
        terminate_procedure ← True // Maximum time reached
        N_{M_max} ← N_M
    end if
    // Outer loop
    for i from 2 to I^outer
        update T
        update d_{N_M}
        s ← s^best
        update N(s, d_{N_M})
        compute I^inner
        // Inner loop
        do inner_loop
    end for
    // Store subproblem optimal configuration
    s_{N_M}^opt ← s^best
    // Check whether next subproblem possible
    estimate t'_{it_{N_M+1}}
    if t'_{it_{N_M+1}} < (t_T - t_c)
        terminate_procedure ← True // Maximum time reached
    end if
end while
// Obtain overall optimal configuration
s^opt ← min_{s ∈ {s_1^opt, s_2^opt, ..., s_{N_{M_max}}^opt}} D(s)
```

Fig. 3. Pseudocode of the proposed HSA procedure. The lines belonging to the inner loop are presented in Fig. 4.

```
for j from 1 to I^inner
    if C(s) equal ∅
        break
    end if
    s' ← random(C(s))
    if D(s') < D(s) or random([0,1]) ≤ exp((D(s) - D(s'))/T)
        s ← s'
        update N(s, d_{N_M})
        if (D(s) < D(s^best))
            s^best ← s
        end if
    end if
    include s in F_{Ω_{N_M}}
end for
```

Fig. 4. Pseudocode of the inner loop of the proposed HSA procedure

*2) Optimization procedure*

The proposed HSA combines both base metaheuristics as summarized in the pseudocode presented in Fig. 3 and Fig. 4. All variables and functions are described in the following subsections.

*a) Temperature, annealing schedule and number of iterations of the outer loop*

The temperature follows a schedule that relies on the time restriction imposed on the optimization problem. The more time the metaheuristic has for solving the problem, the slower the temperature will decrease and the more inner loops there will be. In the proposed procedure, temperature $T$ decreases linearly with the outer loop iteration $i$ in the following fashion:

$$T = \left(\frac{I^{\text{outer}} - i}{I^{\text{outer}} - 1}\right) \cdot T^{\text{init}} \tag{11}$$

where $I^{\text{outer}}$ is the number of iterations of the outer loop, which is computed as:

$$I^{\text{outer}} = \min\left(\left\lfloor\left(\frac{t_{\text{T}} - t_{\text{c}}}{t_{\text{it}}}\right)\right\rfloor, I_{\max}\right) \tag{12}$$

where $t_{\text{T}}$ relies on the imposed time restriction, which will in turn depend on the maximum allowed extra latency regarding the service, $t_{\text{c}}$ is the time that has been consumed so far, which basically includes the time spent solving previous optimization subproblems, and $t_{\text{it}}^{N_M}$ is the time consumed by each outer loop iteration. This value is estimated from measuring the time of the first iteration of the current problem. On the other hand, $I_{\max}$ is set to a constant number, so as to establish a maximum duration to the process. In the case that $I^{\text{outer}}$ is found to be 0 or 1, the process will terminate immediately.

The initial value of the temperature, $T^{\text{init}}$, is set to equal the distortion that is obtained when employing the standard Pro-MPEG COP3 for protecting the data packets in the block (a configuration made up of only one solution, $s_1^{\text{opt}}$). So:

$$T^{\text{init}} = D(s_1^{\text{opt}}) \tag{13}$$

Usually, deriving a good value for $T^{\text{init}}$ is a crucial but hard task, as it requires an approximate knowledge of the cost of the feasible solutions. One common method to acquire this information is through a random sampling of the solution space. However, this process introduces extra latency. Thus, setting it to $D(s^{\text{opt}}_1)$ is a good starting point to the procedure, since this value is in the same order of magnitude as the distortion associated with the configurations in the solution space. In addition, in this way, $T^{\text{init}}$ is always set automatically at the beginning of the process.

Thus, with the proposed values and schedule, three useful results are achieved: to fulfill time requirements, to perform a pseudorandom sampling during the first iterations of the outer loop (so that the whole solution space is inspected to some extent), and to favor a greedy strategy for the last ones.

Finally, every time a subproblem is solved, an estimation of the time that will potentially take an outer iteration of the following subproblem, $t'_{\text{it}_{N_M+1}}$, is carried out to check whether there is still time to complete at least one outer iteration of that subproblem. In the case that $t'_{\text{it}_{N_M+1}}$ is greater than the remaining time, the following subproblem will be posed and started. $t'_{\text{it}_{N_M+1}}$ is computed as follows:

$$t'_{\text{it}_{N_M+1}} = t_{\text{it}_{N_M}\max} \cdot \frac{\widetilde{N}_{\Omega_{N_M+1}}}{\widetilde{N}_{\Omega_{N_M}}} \tag{14}$$

where $t_{\text{it}_{N_M}\max}$ is the maximum time taken by any of the outer iterations of the subproblem just solved, $\widetilde{N}_{\Omega_{N_M+1}}$ is the number of feasible solutions to the following subproblem, and $\widetilde{N}_{\Omega_{N_M}}$ is the number of feasible solutions to the subproblem that has just been solved.

b)   *Neighbors of a solution*

Before discussing the neighborhood of the current solution, let us define the distance between two configurations made up of $N_M$ matrices $s_1 = \{C_1^{s_1}, R_1^{s_1} \dots C_{N_M}^{s_1}, R_{N_M}^{s_1}\}$ and $s_2 = \{C_1^{s_2}, R_1^{s_2} \dots C_{N_M}^{s_2}, R_{N_M}^{s_2}\}$ as:

$$d_{N_M}(s_1, s_2) = \left\{ \sum_{m=1}^{N_M-1} \left[ \left(C_m^{s_1} - C_m^{s_2}\right)^2 + \left(R_m^{s_1} - R_m^{s_2}\right)^2 \right] \right\}^{\frac{1}{2}} \tag{15}$$

In the proposed procedure, the neighbors of a solution $s$, $N(s, d_{N_M})$, are all the feasible solutions in the solution space $\Omega_{N_M}$ within a distance $d'_{N_M}$, i.e.:

$$N(s, d_{N_M}) = \{s' \in \Omega_{N_M} | d_{N_M}(s, s') \leq d'_{N_M}\} \tag{16}$$

where distance $d_{N_M}$ is updated at the beginning of each outer loop iteration according to the same schedule as the temperature (see (11)), in the following fashion:

$$d_{N_M} = \left(\frac{I^{\text{outer}} - i}{I^{\text{outer}} - 1}\right) \cdot d_{N_M}^{\text{init}} \tag{17}$$

where $d_{N_M}^{\text{init}}$ is the value that is used to initialize the distance. This value is set to the potential greatest distance between any two solutions in $\Omega_{N_M}$, which is computed as follows:

$$d_{N_M}^{\text{init}} = \left\{\sum_{m=1}^{N_M-1} [(C_m^{\max} - C_m^{\min})^2 + (R_m^{\max} - R_m^{\min})^2]\right\}^{\frac{1}{2}} \tag{18}$$

where:

$$C_m^{\max} = \max(C_m^{s'} | s' \in \Omega_{N_M}) \quad C_m^{\min} = \min(C_m^{s'} | s' \in \Omega_{N_M})$$

$$R_m^{\max} = \max(R_m^{s'} | s' \in \Omega_{N_M}) \quad R_m^{\min} = \min(R_m^{s'} | s' \in \Omega_{N_M})$$

In this way, at least in the first iteration, all solutions in $\Omega_{N_M}$ can be reached from any other solution.

*c)* *Number of iterations in the inner loop*

The number of iterations in the inner loop, $I^{\text{inner}}$, represents the number of neighbor solutions visited within an iteration of the outer loop. It is set to the maximum value of two options. The first one is a percentage of the number of neighbors of the starting solution $s$ in the current outer loop iteration $i$, $\tau$. The second one is the number of neighbors of $s$ within a distance of $\sqrt{N_M}$, that is, the closest ones. Therefore:

$$I^{\text{inner}} = \max\left(\tau \cdot N(s, d_{N_M}), N(s, \sqrt{N_M})\right) \tag{19}$$

In this way, a sufficient number of configurations in the solution space can be sampled during the first iterations of the outer loop, whereas a minimum number of candidates are ensured for the last ones so as to properly carry out a greedy algorithm.

d)   Memory

Two types of memory structures are used in this procedure, as a legacy of TS: a short-term one, and an intermediate-term one.

The short-term memory structure basically stores all the solutions visited during the course of the process, which are included in a forbidden move list, $F_{\Omega_{N_M}}$, so that they are not tested again.

On the other hand, the intermediate-term memory structure stores just one value, the best solution found during the course of an internal loop, $s_{N_M}^{opt}$. This solution is used as the starting solution for the following inner loop, as a means to intensify the search of solutions in promising regions of the solution space.

Long-term memory structures are not necessary, as a type of diversification is actually performed during the first outer iterations of the procedure, thanks to the management of neighborhood structures.

e)   Candidate solutions

If $s$ is the current solution, the candidate solutions $C(s)$ are all the solutions in the solution space the system can move to from $s$, that is, those ones that are neighbors of $s$ and have not been visited yet. This is expressed next:

$$C(s) = \left\{ s' \in N(s, d_{N_M}) \wedge s' \notin F_{\Omega_{N_M}} \right\} \qquad (20)$$

Table 1. System characterizing variables

| $R$ (Mbps) | $T_L$ (s) | $T_R$ (s) | $r_{FEC}$ | $N_P$ | $N_{FEC}$ |
|---|---|---|---|---|---|
| 4 | 1 | 0.5 | 10/11 | 185 | 19 |
| 4 | 1 | 0.5 | 5/6 | 185 | 37 |
| 4 | 0.2 | 0.1 | 10/11 | 37 | 4 |
| 4 | 0.2 | 0.1 | 5/6 | 37 | 7 |
| 8 | 1 | 0.5 | 10/11 | 370 | 37 |
| 8 | 1 | 0.5 | 5/6 | 370 | 74 |
| 8 | 0.2 | 0.1 | 10/11 | 74 | 7 |
| 8 | 0.2 | 0.1 | 5/6 | 74 | 15 |
| 12 | 1 | 0.5 | 10/11 | 556 | 56 |
| 12 | 1 | 0.5 | 5/6 | 556 | 111 |
| 12 | 0.2 | 0.1 | 10/11 | 111 | 11 |
| 12 | 0.2 | 0.1 | 5/6 | 111 | 22 |

The procedure parameters $N_P$ and $N_{FEC}$ are estimated from the other three parameters.

**Table 2 Results of the first set of experiments**

| $N_P$ | $N_{FEC}$ | $N_M$ | $N_{\Omega_{N_M}}$ | $\widetilde{N}_{\Omega_{N_M}}$ | Exhaustive search (not restricted) | | Exhaustive search (restricted) | | HSA (restricted) | |
|---|---|---|---|---|---|---|---|---|---|---|
| | | | | | Time (s) | PSNR (dB) | Time (s) | PSNR (dB) | Time (s) | PSNR (dB) |
| 185 | 19 | 2 | 590 | 85 | 1.03E-02 | 28.79 | 7.82E-04 | 28.79 | 6.69E-04 | 28.79 |
| 185 | 37 | 2 | 638 | 90 | 6.94E-03 | 33.0 | 4.74E-04 | 33.0 | 4.18E-04 | 33.0 |
| 37 | 4 | 2 | 63 | 18 | 2.84E-04 | 28.46 | 7.74E-05 | 28.46 | 1.13E-04 | 28.46 |
| 37 | 7 | 2 | 79 | 15 | 2.06E-04 | 33.21 | 7.05E-05 | 33.21 | 8.84E-05 | 33.2 |
| 185 | 19 | 3 | 154921 | 3887 | 1.64 | 28.87 | 5.07E-02 | 28.87 | 6.92E-03 | 28.86 |
| 185 | 37 | 3 | 191941 | 3999 | 1.62 | 33.14 | 3.76E-02 | 33.14 | 4.92E-03 | 33.11 |
| 37 | 4 | 3 | 1207 | 81 | 6.5E-03 | 28.53 | 4.28E-04 | 28.53 | 4.22E-04 | 28.53 |
| 37 | 7 | 3 | 2384 | 121 | 9.31E-03 | 33.32 | 5.56E-04 | 33.32 | 4.9E-04 | 33.32 |
| 185 | 19 | 4 | 24045652 | 93752 | 311.91 | 28.9 | 1.33 | 28.9 | 6.22E-02 | 28.89 |
| 185 | 37 | 4 | 35985286 | 106826 | 380.24 | 33.17 | 1.06 | 33.17 | 6.23E-02 | 33.14 |
| 37 | 4 | 4 | 7140 | 378 | 3.22E-02 | 28.53 | 2.15E-03 | 28.53 | 2.19E-03 | 28.53 |
| 37 | 7 | 4 | 36227 | 427 | 2.15E-02 | 33.34 | 2.49E-03 | 33.34 | 2.59E-03 | 33.34 |

## VI. EXPERIMENTS AND RESULTS

With the aim of evaluating the effects of both the solution space reduction and the proposed HSA procedure, an exhaustive set of experiments were carried out. In these tests, three video sequences of dissimilar bitrate, $R$, were considered: $R_1 = 4$ Mbps (Seq1), $R_2 = 8$ Mbps (Seq2) and $R_3 = 12$ Mbps (Seq3). Moreover, two values were used to limit FEC latency, $T_L = 200$ ms and 1 s, which correspond approximately to typical constraints imposed on live event and non-live event broadcasts, respectively. As XOR operations are considered negligible in terms of time, roughly the whole delay introduced by standard Pro-MPEG COP3 codes is due to the time that the receiver needs to wait for all the source and repair packets that belong to the same matrix, so that decoding operations can be carried out [3], [37]. However, in the case of the UEP version, the procedure to obtain the optimal configuration also introduces some extra delay. So, with the aim of always complying with the imposed time restriction, it becomes necessary to split the time that is devoted to FEC-related operations between the receiver, $T_R$, and the transmitter, $T_T$, so that there is time to wait for the packets, as in the standard case, and the optimization procedure can be carried out. In these experiments, half of the FEC time is used by the receiver and the other half by the transmitter. So, approximately only half of the data packets are handled at each step of the UEP version, when compared to the standard one. This fact is detrimental to the obtained results, as, generally, the efficiency of FEC codes (and Pro-MPEG COP3 codes are no exception) increases with the number of data packets that are handled together. However, the fact of exploiting the unequal importance of video packet not only counteracts this drawback in most scenarios, but delivers better results in terms of overall distortion, as will be shown. Finally, two different channel code rates, $r_{FEC}$, were employed: $r_{FEC_1}$ = 10/11 (10% overhead) and $r_{FEC_2}$ = 5/6 (20% overhead). Table 1 shows these system characterizing variables, and an approximation of the values of parameters $N_P$ and $N_{FEC}$, which entirely rely on them.

Two sets of experiments were carried out to assess three different aspects of the proposal: the solution space reduction, the performance of the proposed HSA procedure, and the adaptive capacity of the algorithm to the imposed time restrictions and the specifics of the scenario at hand, jointly with its effectiveness under those conditions. The first two aspects were assessed in terms of processing time (with a 2-core CPU clocked at 3 GHz with 12 GiB RAM), and resulting distortion in terms of peak-to-noise rate (PSNR). The latter aspects were evaluated in terms of processing time (to validate adaptability; with the same CPU) and overall distortion, again in terms of PSNR. For all the experiments, a fairly simple distortion model was employed to obtain the importance of source packets: the distortion associated with the transmission of a given data packets equals the number of packets in the current Group of Pictures (GOP) that depend on it for decoding. Despite constituting a rather simple distortion model, it suitably reflects two main aspects of how video streams behaves when impacted by packet losses: inter-frame error propagation and intra-frame desynchronization, and provides a relative hierarchization of data packets that suffices for properly illustrating the key factors that impact on the performance of the strategy. Nevertheless, the proposed strategy can surely benefit from using more realistic approaches, as they commonly provide greater differences of relevance between data packets than those obtained with the considered model. This stronger hierarchization can noticeably increase the effectiveness of the proposed UEP, as predicted in [36].

In the first set of experiments, the following scenario was set: Seq1, PLR = 1.0E-2, ABL = 1 ms. Furthermore, the number of iterations of the outer loop, $I^{outer}$, was set to 10 (no time restrictions were imposed), and the subproblems dealing with 2 to 4 matrices were considered. Simulations were run 100 times and the obtained results averaged afterwards.

As can be observed in Table 2, the conditions introduced to restrict the solution space absolutely make the number of configurations go down, whereas barely affecting the result. Thus, the optimization procedure starts from a more advantageous situation. Moreover, it can also be seen that the proposed procedure is able to find sufficiently good solutions in a significantly shorter period of time, at the expense of a very small error, almost negligible in a logarithmic scale, which slightly increases with the size of the solution space. In particular, the larger the solution space is, the greater it is the gain in terms of time.

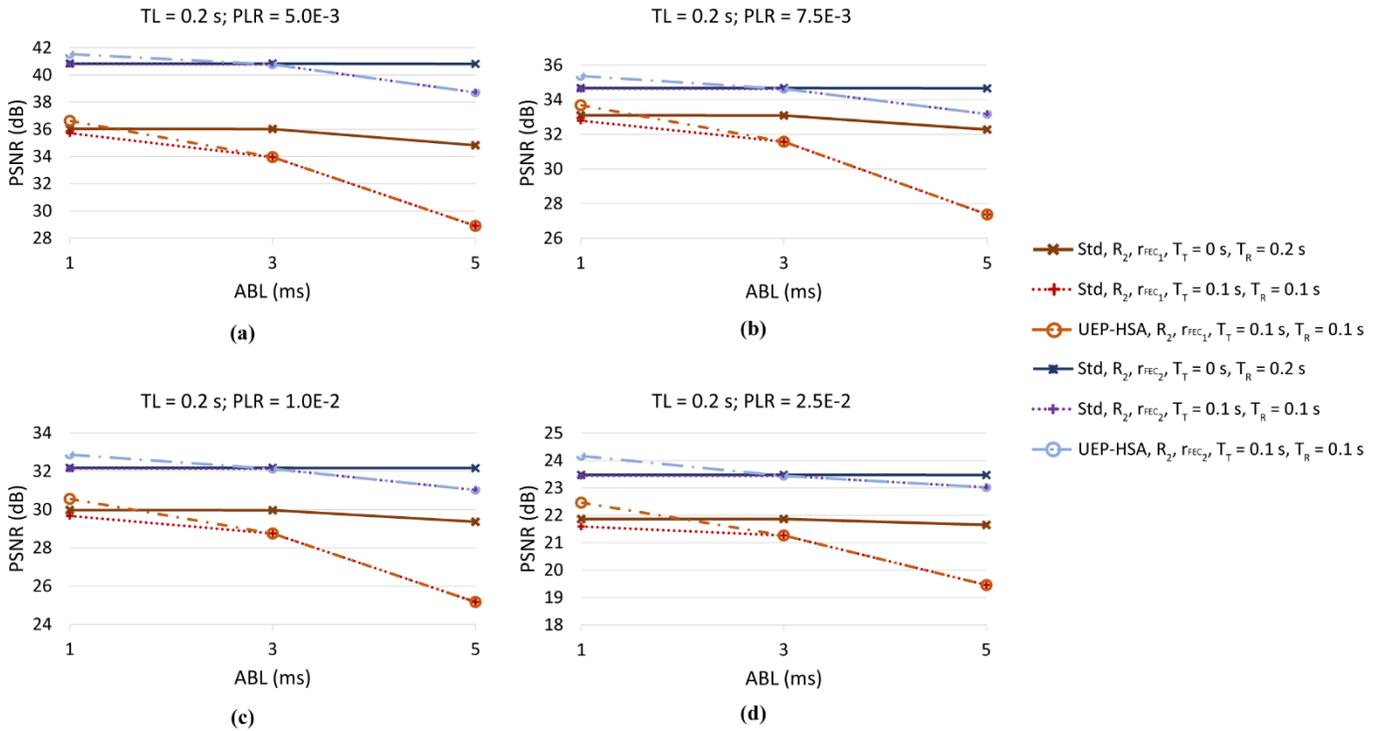

Fig. 5. Results of the second set of experiments in terms of distortion when the whole imposed FEC latency equals 200 ms. The results of the standard codes using the whole imposed FEC latency (200 ms) are presented using solid lines. The results of the standard codes using half of the imposed FEC latency (100 ms) are presented using dotted lines in different colors. The results of the UEP version of the codes optimized through the proposed HSA procedure are presented using dashed lines in different colors.

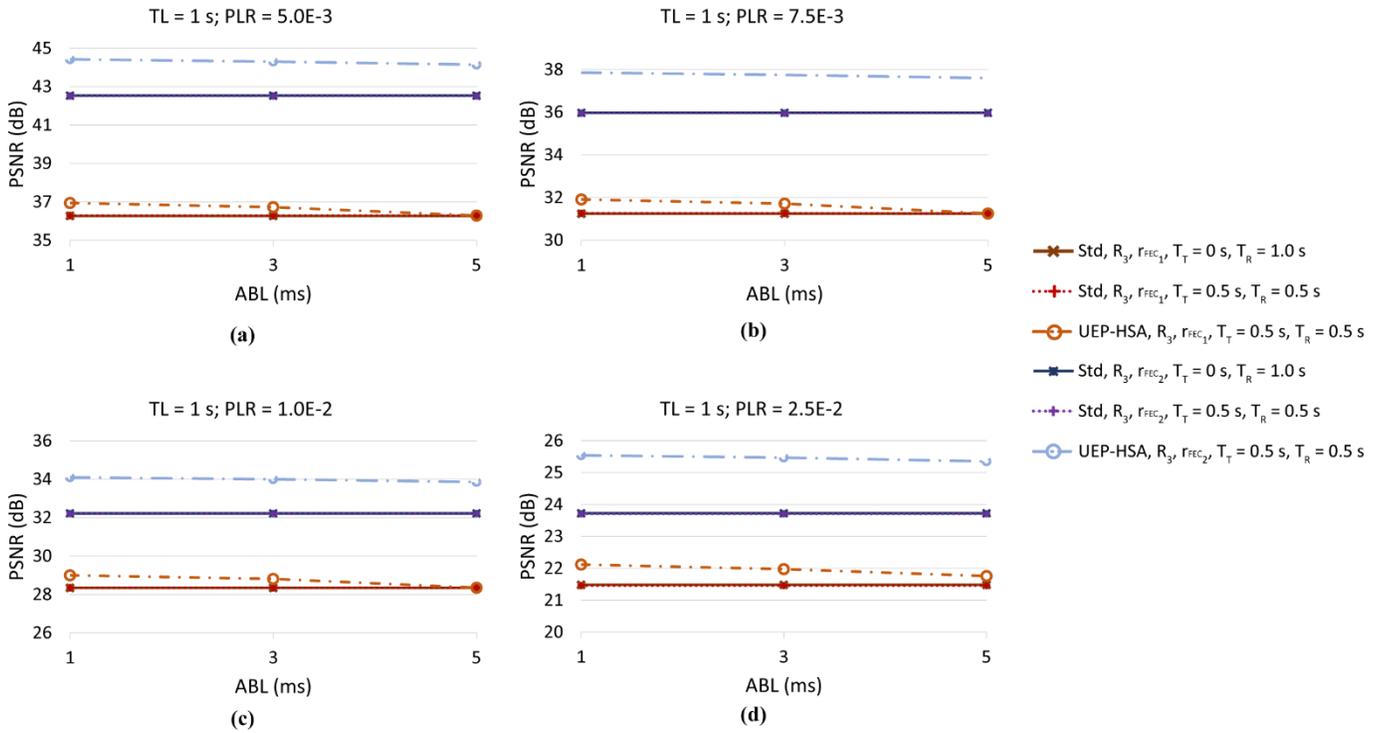

**Fig. 6.** Results of the second set of experiments in terms of distortion when the whole imposed FEC latency equals 1 s. The results of the standard codes using the whole imposed FEC latency (1 s) are presented using solid lines. The results of the standard codes using half of the imposed FEC latency (500 ms) are presented using dotted lines in different colors. The results of the UEP version of the codes optimized through the proposed HSA procedure are presented using dashed lines in different colors.

The second set of experiments were aimed at validating the proposed strategy when strict time restrictions apply. In these experiments, Seq2 and Seq3 were employed, and $I^{outer}$ was set to vary according to (12). Moreover, a simplified Gilbert-Elliot model was used to model the behavior of the channel, which is described through two parameters: the packet loss rate (PLR) and the average burst length (ABL). In these simulations, four different PLR and three ABL values were considered. The ABL values reflect the average length of the periods where consecutively transmitted data packets are dropped during transmission, and are given in terms of time, which translate into different numbers of consecutive loss packets, regarding the bitrate of the video stream. The PLR values have been chosen in the region where standard Pro-MPEG COP3 codes start presenting a weaker performance, which thus correspond to our points of interest. Again, simulations were run 100 times and the obtained results averaged afterwards.

The results of these experiments regarding the introduced overall distortion are shown in Fig. 5 and Fig. 6. These plots present the distortion introduced when using the standard codes using the whole imposed maximum FEC time (which is used as reference), and the distortion introduced when using the UEP version of the codes optimized through the proposed metaheuristic, where half of the imposed maximum FEC latency is devoted to carry out the algorithm and the other half at the receiver. Results presenting the distortion value obtained when using the standard codes employing the whole available FEC time and that obtained when using only half of it are also included for comparison. In particular, this comparison allows on the one hand to capture the detrimental effect mentioned before, and, on the other hand, to directly measure the benefits of employing the UEP version of the codes with respect to the standard when employing the same protection block size. As already mentioned, distortion values are provided in terms of PSNR. The results corresponding to $T_L$ = 200 ms are depicted in Fig. 5, whereas the ones corresponding to $T_L$ = 1 s are presented in Fig. 6. As the results on the two sequences show very similar trends, due to space limitations, only the ones related to sequence Seq2 are depicted for the scenario with $T_L$ = 200 ms, and only the ones related to sequence Seq3 are provided for the scenario with $T_L$ = 1 s.

As can be seen, the effectiveness of the proposed procedure increases with the capacity to generate configurations that are capable of decorrelating channel error, that is, with the number of data and repair packets per protection block, and inversely with the length of error bursts. Specifically, the greater it is the ABL of the communication channel, the more columns are required to provide suitable interleaving and so decorrelate error. This means that in the scenarios where it is not possible to generate configurations where matrices have a sufficient number of columns to cope with channel's ABL, that is, where $T_L$ (and therefore the number of data and repair packets per block) is low in comparison with the existing ABL value, the results obtained when applying the adopted UEP scheme cannot outperform those obtained with the standard codes, as none of the feasible configurations can properly deal with burst errors (see Fig. 5). On the other hand, if $T_L$ is high enough compared to channel's ABL, there can be found configurations that can correctly cope with error bursts. In this case, results are less dependent with the particular ABL value, as can be extracted from the results of the experiments where $T_L$ = 1 s (see Fig. 6), where figures are flatter. $T_L$ is in this case sufficiently high in comparison with the considered error burst lengths: 1, 3 and 5 ms. Furthermore, if the channel's PLR value increases, the likelihood of two or more error bursts impacting the same matrix and moreover affecting packets in the same column increases as well. In this situation, the proposed UEP strategy is capable of reducing the number of rows of the matrices where the most important data packets are to be arranged (and so decrease the code rate and therefore increase the protection capability of the code), and so make that likelihood lower for the more important packets, thus obtaining a better overall distortion results than those of the standard strategy, even for low $T_L$ values. This circumstance can be seen in both Fig. 5 and Fig. 6, where the difference in dBs between the UEP and the standard versions remains fairly constant, despite the PLR increase, which indicates that the relative difference increases. Obviously, in the scenarios where the PLR is particularly high, those extra dBs might not make a real difference to the user (e.g. if PLR = 2.5E-2). However, there exist a considerably wide interval of PLR values where this disparity will make the difference between an acceptable and an unacceptable viewing experience.

So, with good-enough conditions, the UEP codes optimized using HSA can easily obtain a PSNR increase of up to several dBs with respect to the standard approach, depending on the scenario. Moreover, as mentioned above, the use of distortion models of the packet stream that more accurately capture the unequal importance of data packets will lead with high probability to deliver even better results. Finally, it is worth noticing that results also show that the gap between applying the standard codes using the whole FEC period and using only half of it decreases with the capacity of being able to decorrelate error. However, the latter never outperform the former, unless the UEP version is applied.

Table 3. Results of the second set of experiments in terms of processing time

| $R$ (Mbps) | $r_{FEC}$ | $T_L$ (s) | $T_T$ (s) | Time (s) | | | $N_{M_{max}}$ |
|---|---|---|---|---|---|---|---|
| | | | | Mean | Maximum | Variance | |
| 8 | 10/11 | 0.2 | 0.1 | 8.84E-02 | 9.83E-02 | 1.48E-04 | 5.96 |
| 8 | 5/6 | 0.2 | 0.1 | 9.12E-02 | 9.97E-02 | 2.42E-04 | 5.67 |
| 8 | 10/11 | 1 | 0.5 | 3.74E-01 | 4.85E-01 | 1.95E-02 | 4.95 |
| 8 | 5/6 | 1 | 0.5 | 3.93E-01 | 4.86E-01 | 1.42E-02 | 4.5 |
| 12 | 10/11 | 0.2 | 0.1 | 9.30E-02 | 9.91E-02 | 1.52E-04 | 3.85 |
| 12 | 5/6 | 0.2 | 0.1 | 6.79E-02 | 9.95E-02 | 9.75E-04 | 3.91 |
| 12 | 10/11 | 1 | 0.5 | 4.24E-02 | 6.29E-02 | 5.09E-05 | 3 |
| 12 | 5/6 | 1 | 0.5 | 3.58E-02 | 5.66E-02 | 4.58E-05 | 3 |

The performance of the procedure is assessed in terms of processing time (mean, maximum and variance), and average number of subproblems finally posed per step along the sequence.

To verify the adaptability to these constraints and the performance of the scheme, both the average time spent on carrying out the optimization (all mean, maximum and variance values are included) and the average number of subproblems that it was possible to pose, given those time constraints. Finally, in the same way as before, simulations have been run 100 times and the results averaged afterwards.

As can be seen in Table 3, the algorithm is able to adapt on the fly to strict time conditions and comply with them. In most cases, the procedure uses the most part of the available FEC time. Nevertheless, in some circumstances, like that in which Seq3 is protected allowing a FEC latency of 1 s, the time that is finally spent in applying the metaheuristic is not close to the limit. This means that at the end of one of the subproblems (in this example, the third one), the time that one outer iteration of the following subproblem is estimated to last exceeds the time that is left (see (14)). Therefore, the procedure ends, even if there is some FEC time left. Finally, as expected, the average number of subproblems that can be posed and started per block along the sequence increases inversely with the number of feasible solutions in the solution space, that is, with the number of data and repair packets per FEC block.

Table 4. Example of the performance of the proposed strategy.

| $N_M$ | Final solution | Distortion (normalized) | Resulting error rate per matrix |
|---|---|---|---|
| 1 | $[15\text{x}5]_1$ | 1 | $[4.9\text{E-}4]_1$ |
| 2 | $[13\text{x}4]_1$ $[2\text{x}11]_2$ | 0.84 | $[3.94\text{E-}4]_1$ $[10.47\text{E-}4]_2$ |
| 3 | $[9\text{x}3]_1$, $[5\text{x}6]_2$, $[1\text{x}17]_3$ | 0.78 | $[2.97\text{E-}4]_1$, $[5.85\text{E-}4]_2$, $[15.71\text{E-}4]_3$ |
| 4 | $[7\text{x}3]_1$, $[5\text{x}4]_2$, $[2\text{x}7]_3$, $[1\text{x}19]_4$ | 0.77 | $[2.97\text{E-}4]_1$, $[3.94\text{E-}4]_2$, $[6.79\text{E-}4]_3$, $[17.38\text{E-}4]_4$ |
| 5 | $[7\text{x}3]_1$, $[4\text{x}4]_2$, $[2\text{x}6]_3$, $[1\text{x}9]_4$, $[1\text{x}16]_5$ | 0.76 | $[2.97\text{E-}4]_1$, $[3.94\text{E-}4]_2$, $[5.85\text{E-}4]_3$, $[8.65\text{E-}4]_4$, $[14.85\text{E-}4]_5$ |
| 6 | $[8\text{x}3]_1$, $[2\text{x}4]_2$, $[2\text{x}5]_3$, $[1\text{x}6]_4$, $[1\text{x}9]_5$, $[1\text{x}17]_6$ | 0.78 | $[2.97\text{E-}4]_1$, $[3.94\text{E-}4]_2$, $[4.9\text{E-}4]_3$, $[5.85\text{E-}4]_4$, $[8.65\text{E-}4]_5$, $[15.71\text{E-}4]_6$ |
| 7 | $[4\text{x}3]_1$, $[3\text{x}3]_2$, $[3\text{x}4]_3$, $[2\text{x}5]_4$, $[1\text{x}6]_5$, $[1\text{x}8]_6$, $[1\text{x}17]_7$ | 0.81 | $[2.97\text{E-}4]_1$, $[2.97\text{E-}4]_2$, $[3.94\text{E-}4]_3$, $[4.9\text{E-}4]_4$, $[5.8\text{E-}4]_5$, $[7.73\text{E-}4]_6$, $[15.7\text{E-}4]_7$ |

$[C_k x R_k]_k$ means that the $k^{\text{th}}$ matrix in the configuration has $R_k$ rows and $C_k$ columns. Additionally, $[P_k]_k$ is the resulting error rate of the data packets protected through matrix $k$.

Finally, we assess the adaptability of the scheme to the specifics of the scenario. The strategy proposed in this paper potentially provides a different solution for each group of consecutive data packets along the stream, mostly depending on the relative importance of the data packets included in each protection block. This is the reason why unique overall solutions are not provided as a result of the transmission of an encoded sequence under certain conditions. However, an illustrative example is included to fully clarify and characterize the performance of the protection strategy.

This example considers the protection of the first data packets in the corresponding stream. The scenario taken into account is as follows: $T_L$ = 200 ms, Seq2, $r_{\text{FEC}_2}$, PLR = 1.0E-2, ABL = 1 ms. So, the first $N_P$ = 74 data packets in the stream are arranged for protection in the first block and $N_{\text{FEC}}$ = 15 are to be generated. First, the standard strategy is considered. After this, subproblems are posed sequentially until the available time is over. Table 4 shows some important results of each subproblem: final configuration, relative distortion with respect to the standard configuration, and resulting error rate per matrix.

In this case, the presented configuration with five matrices is the best solution that could be found, and so it is the one used to arrange data packets and generated packets. The most important 21 data packets are arranged in the first matrix and seven parity packets are generated. The following 16 most important packets are arranged in the second one and four repair packets are generated. And so on. The resulting error rate per matrix considers the probability of not receiving a packet and also not being able to recover it. The presented value is the average of all the data packets in each matrix.

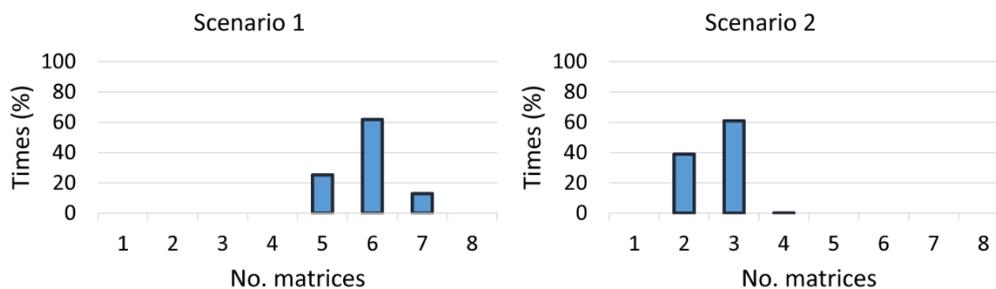

Fig. 7. Histograms showing the percentage of selection of configurations that are made up of each number of matrices.

Finally, with the aim of showing the adaptive nature of our proposal, we present, for this same scenario and a second one with quite different characteristics ($T_L$ = 1 s, Seq3, $r_{FEC_2}$, PLR = 7.5E-5, ABL = 3 ms), a histogram that shows the percentage of times a configuration of each number of matrices is selected as the best option during the transmission of the whole sequence. The first video stream consists of 641580 packets, that is, 8670 blocks of $N_P$ = 74 data packets, whereas the second one is made up of 639400 packets, i.e., 1150 blocks of $N_P$ = 556 data packets. Such histograms are depicted in Fig. 7.

It can be observed that, in the first scenario, the protection strategy opts more often for configurations made up of six matrices, followed by configurations of five and seven. In the case of the second scenario, the protection technique almost always selects configurations of two or three matrices, with greater preference for the latter. The selection of a given configuration is, as already seen, the result of a number of elements: available time, channel behavior, characteristics of the transmitted video, etc. Scenarios with very strict time limitations and large block sizes, and/or where packets are rather similar in terms of the distortion that they might cause if they are lost (e.g. encoded sequences made up of short closed GOPs) tend to select configurations of few matrices. On the other hand, if there is enough time to pose and solve the necessary subproblems, and the distortion associated with the different packets is sufficiently unequal (e.g. long GOPs with hierarchical B-frames), the protection mechanism tends to select configuration with more matrices.

VII. CONCLUSION

In this paper, a metaheuristic that results from the hybridization of two powerful optimization techniques, simulated annealing and tabu search, has been presented. The proposed optimization method is situated in the context of the UEP mechanisms that aim at smartly protecting audio and video transmission over IP networks when providing very time-sensitive services. The considered code, a UEP version of the Pro-MPEG COP3 codes, allows the utilization of a potentially very large number of protection configurations. So, to fulfill the imposed timing constraints, and still be able to find near-optimal protection configurations, SA is adapted to the specifics of the considered scenario.

The proposed method consists of two steps. In the first one, the solution space is significantly reduced through imposing UEP-aware restrictions. In the second one, SA's core procedure is modified to take into account imposed time restrictions and explore the solution space considering a proposed definition of distance between configurations that allows a periodic update of neighbor structures. This result is also in part achieved thanks to the incorporation of memory structures from TS.

A wide number of simulations have been carried out and grouped in two different experiments, to assess the performance of the proposed method. The first set of experiments validates the reduction of the solution space and the performance of the proposed procedure. The second one verifies that the HSA method is able to invariably fulfill real-time constraints, and simultaneously continually adapt to the particularities of the scenario at hand so as to find sufficiently good protection configurations that allow the UEP Pro-MPEG COP3 codes to clearly outperform the standard codes in a wide range of scenarios.

## REFERENCES

[1] M. Luby, T. Stockhammer, and M. Watson, "IPTV systems, standards and architectures: Part II - Application Layer FEC in IPTV services," *IEEE Communications Magazine.*, vol. 46, no. 5, pp. 94–101, May 2008.

[2] F. Battisti, M. Carli, E. Mammi, and A. Neri, "A study on the impact of AL-FEC techniques on TV over IP Quality of Experience," *EURASIP Journal on Advances in Signal Processing*, vol. 2011, no. 1, p. 86, 2011.

[3] ETSI TS 102 542-3-2, "Digital Video Broadcasting (DVB): Guidelines for the implementation of DVB-IP Phase 1 specifications. Part 3: Error Recovery. Subpart 2: Application Layer FEC," May 2011.

[4] A. Nafaa, T. Taleb, and L. Murphy, "Forward Error Correction Strategies for Media Streaming over Wireless Networks," *IEEE Communications Magazine*, vol. 46, no. 1, pp. 72–79, Jan. 2008.

[5] Y. Huo, C. Hellge, T. Wiegand, and L. Hanzo. "A tutorial and review on inter-layer FEC coded layered video streaming," *IEEE Communications Surveys & Tutorials*, vol. 17, no 2, p. 1166-1207, 2015.

[6] C. Hellge, D. Gomez-Barquero, T. Schierl, and T. Wiegand, "Layer-aware forward error correction for mobile broadcast of layered media," *IEEE Transactions on Multimedia*, vol. 13 no. 3, p. 551–562, Jun. 2011.

[7] C. Díaz, J. Cabrera, F. Jaureguizar, and N. García, "A video-aware FEC-based unequal loss protection system for video streaming over RTP," *IEEE Transactions on Consumer Electronics*, vol. 57, no. 2, pp. 523-531, May 2011.

[8] T. Cheng, K. Peng, F. Yang, J. Song, and Y. Zhixing, " A Near-Capacity MIMO Coded Modulation Scheme for Digital Terrestrial Television Broadcasting," *IEEE Transactions on Broadcasting*, vol. 61, no. 3, pp. 367-375, Sep. 2015.

[9] C. Ye, G. Ozcan, M. C. Gursoy, and S. Velipasalar, "Multimedia Transmission Over Cognitive Radio Channels Under Sensing Uncertainty," *IEEE Transactions on Signal Processing*, vol. 64, no. 3, pp. 726-741, Feb. 2016.

[10] S. K. Srinivasan, J. Vahabzadeh-Hagh, and M. Reisslein, " The Effects of Priority Levels and Buffering on the Statistical Multiplexing of Single-Layer H.264/AVC and SVC Encoded Video Streams," *IEEE Transactions on Broadcasting*, vol. 56, no. 3, pp. 281-287, Sep. 2010.

[11] C. Zhou, C. W. Lin, X. Zhang, and Z. Guo, "A Novel JSCC Scheme for UEP-Based Scalable Video Transmission Over MIMO Systems," *IEEE Transactions on Circuits and Systems for Video Technology*," vol. 25, no. 6, pp. 1002-1015, Jun. 2015.

[12] S. Sandberg, and N. von Deetzen, "Design of bandwidth-efficient unequal error protection LDPC codes*," IEEE Transactions on Communications,* vol. 58, no. 3, pp. 802-811, Mar. 2010.

[13] A. Vosoughi, P. C. Cosman, and L. B. Milstein, "Joint Source-Channel Coding and Unequal Error Protection for Video Plus Depth," *IEEE Signal Processing Letters*, vol. 22, no. 1, pp. 31-34, Jan. 2015.

[14] Y. C. Chang, S. W. Lee, and R. Komiya, "A low complexity hierarchical QAM symbol bits allocation algorithm for unequal error protection of wireless video transmission," *IEEE Transactions on Consumer Electronics*, vol. 55, no. 3, pp. 1089-1097, Aug. 2009.

[15] H. Ha, and C. Yim, "Layer-weighted unequal error protection for scalable video coding extension of H.264/AVC," *IEEE Transactions on Consumer Electronics*, vol. 54, no. 2, pp. 736-744, May 2008.

[16] C. Blum, and A. Roli, "Metaheuristics in combinatorial optimization: Overview and conceptual comparison," *ACM Computing Surveys*, 2003, vol. 35, no 3, p. 268-308.

[17] Y. Wu, S. Kumar, F. Hu, Y. Zhu, and J. D. Matyjas, "Cross-Layer Forward Error Correction Scheme Using Raptor and RCPC Codes for Prioritized Video Transmission Over Wireless Channels," *IEEE Transactions on Circuits and Systems for Video Technology*, vol. 24, no. 6, pp. 1047-1060, Jun. 2014.

[18] T. Fang, and L-P. Chau, "GOP-based channel rate allocation using genetic algorithm for scalable video streaming over error-prone networks," *IEEE Transactions on Image Processing*, vol. 15, no. 6, pp. 1323-1330, Jun. 2006.

[19] V. M. Stankovic, R. Hamzaoui, Y. Charfi, and Z. Xiong, "Real-time unequal error protection algorithms for progressive image transmission," *IEEE Journal on Selected Areas in Communications*, vol. 21, no. 10, pp. 1526-1535, Dec. 2003.

[20] L. Cao, "On the Unequal Error Protection for Progressive Image Transmission," *IEEE Transactions on Image Processing*, vol. 16, no. 9, pp. 2384-2388, Sep. 2007.

[21] M. Gendreau and J.-Y. Potvin, editors. Handbook of metaheuristics, Springer; 2010.

[22] C. R. Reeves, "Genetic Algorithms," in Handbook of Metaheuristics, Springer, 2010.

[23] M. Dorigo and T. Stützle, "Ant Colony Optimization: Overview and Recent Advances," in Handbook of Metaheuristics, Springer, 2010.

[24] R. Poli, J. Kennedy and T. Blackwell, "Particle swarm optimization. An overview". *Swarm Intelligence*, vol. 1, no. 1, pp. 33—57, Jun. 2007.

[25] A. G. Nikolaev S. H. Jacobson "Simulated Annealing," in Handbook of Metaheuristics, Springer, 2010.

[26] M. Gendreau and J.-Y. Potvin, "Tabu Seach," in Handbook of Metaheuristics, Springer, 2010.

[27] C. Voudouris, E. P.K. Tsang, A. Alsheddy, "Guided Local Seach," in Handbook of Metaheuristics, Springer, 2010.

[28] R. S. Sexton, R. E. Dorsey and, J. D. Johnson,"Optimization of neural networks: A comparative analysis of the genetic algorithm and simulated annealing," *European Journal of Operational Research*, vol. 114, 589—601, 1999.

[29] B. Dalanezi Mori, H. Fiori de Castro and K. L. Cavalca "Development of hybrid algorithm based on simulated annealing and genetic algorithm to reliability redundancy optimization," *International Journal of Quality & Reliability Management*, vol. 24, no. 9, pp. 972—987, 2007.

[30] S. Kannan, S. M. R. Slochanal, and N. P. Padhy "Application and Comparison of Metaheuristic Techniques to Generation Expansion Planning Problem," *IEEE Transactions on Power Systems*, vol. 20, no 1, Feb. 2005.

[31] M. Garcia-Lozano, M. Lema, S. Ruiz, and F. Minerva, "Metaheuristic Procedure to Optimize Transmission Delays in DVB-T Single Frequency Networks," *IEEE Transactions on Broadcasting*, vol. 57, no. 4, pp. 876-887, Dec. 2011.

[32] D. Cheng, J. Rao, C. Jiang, and X. Zhou, "Elastic Power-Aware Resource Provisioning of Heterogeneous Workloads in Self-Sustainable Datacenters," *IEEE Transactions on Computers*, vol. 65, no. 2, pp. 508-521, Feb. 2016.

[33] M. F. Uddin, H. M. K. AlAzemi, and C. Assi, "Optimal Flexible Spectrum Access in Wireless Networks with Software Defined Radios," *IEEE Transactions on Wireless Communications*, vol. 10, no. 1, pp. 314-324, Jan. 2011.

[34] Y. A. Katsigiannis, P. S. Georgilakis and E. S. Karapidakis, "Hybrid Simulated Annealing–Tabu Search Method for Optimal Sizing of Autonomous Power Systems With Renewables," *IEEE Transactions on Sustainable Energy*, vol. 3, no. 3, pp. 330-338, Jul. 2012.

[35] K. Lenin, B. R. Reddy and M. Suryakalavathi "Hybrid Tabu search-simulated annealing method to solve optimal reactive power problem," *International Journal of Electrical Power & Energy Systems*, vol. 82, pp. 87–91, Nov, 2016.

[36] C. Díaz, J. Cabrera, F. Jaureguizar, and N. García, "An Extension to the Pro-MPEG COP3 Codes for Unequal Error Protection of Real-time Video Transmission," *IEEE Int. Conf. Image Processing (ICIP'15)*, Sep. 2015.

[37] Pro-MPEG Forum Code of Practice 3 release 2, "Transmission of professional MPEG-2 transport streams over IP networks," Jul. 2004.

[38] T. Wiegand, G. J. Sullivan, G. Bjontegaard, and A. Luthra. "Overview of the H.264/AVC video coding standard," *IEEE Transactions on Circuits and Systems for Video Technology*, vol. 13, no. 7, pp. 560–576, Jul. 2003.

[39] G. J. Sullivan, J-R. Ohm, W.-J. Han, and T. Wiegand, "Overview of the High Efficiency Video Coding (HEVC) Standard," *IEEE Transactions on Circuits and Systems for Video Technology*, vol. 22, no. 12, pp. 1649--1668, Dec. 1012.

[40] H. Schulzrinne, S. Casner, R. Frederick, and V. Jacobson, "RTP: A Transport Protocol for Real-Time Applications Status," RFC 3550, 2003.

[41] ISO/IEC, "High efficiency coding and media delivery in heterogeneous environments -- Part 1: MPEG media transport (MMT)", ISO/IEC 23008-1, 2014.

[42] ISO/IEC, "MPEG-2 Part 1, Systems," ISO/IEC 13818-1, 1995.

[43] SMPTE ST 2022-1, "Forward error correction for real-time video/audio transport over IP networks," May 2007.

[44] Digital Video Broadcasting (DVB): "Transport of MPEG-2 TS-based DVB services over IP-based networks," ETSI TS 102 034, Aug. 2009.

[45] Digital Video Broadcasting (DVB); "Upper Layer FEC for DVB Systems", ETSI TR 102 993, Feb. 2011.

[46] A. Shokrollahi, "Raptor codes," *Trans. on Information Theory*, vol. 25, no. 6, pp. 2511-2567, Jun. 2006.

[47] A. Begen "RTP Payload Format for 1-D Interleaved Parity Forward Error Correction (FEC)," RFC 6015, Oct. 2010.

[48] S. Kirkpatrick, C. D. Gelatt, Jr., and M. P. Vecchi, "Optimization by Simulated Annealing*,*" *Science,* vol. 220, no. 4598, pp. 671–680, May 1983.

[49] D. Zhang, Y. Liu, R. M'Hallah, and S. C. H. Leung, "A simulated annealing with a new neighborhood structure based algorithm for high school timetabling problems," *European Journal of Operational Research*, vol. 203, no. 3, pp. 550-558, Jun. 2010.

[50] X. Yao, "Dynamic neighbourhood size in simulated annealing," *International Joint Conference on Neural Networks (IJCNN'92)*, vol. 1, pp. 411-416, Nov. 1992.

[51] F. Glover, "Future paths for integer programming and links to artificial intelligence," *Computers and Operations Research*, vol. 13, no. 5, pp. 533–549, May 1986.